\documentclass[aps,prl,floatfix,twocolumn,notitlepage,superscriptaddress,10pt]{revtex4-2}
\usepackage{xcolor}
\colorlet{RED}{red}
\usepackage{grffile}
\usepackage{amsmath, amsthm, amssymb, bbold}
\usepackage[normalem]{ulem}
\usepackage{cancel}
\usepackage{bm}
\usepackage{microtype}
\usepackage{hyperref}
\usepackage{setspace}
\usepackage{grffile}
\hypersetup{colorlinks,linkcolor=blue,urlcolor=blue,citecolor=blue}
\usepackage{amsmath}    
\usepackage{amssymb}
\usepackage{graphicx}   
\usepackage{verbatim}   
\usepackage{color}      
\usepackage{microtype}
\usepackage[normalem]{ulem}
\usepackage{natbib}
\usepackage{enumitem} 
\usepackage{dsfont} 
\usepackage{hyperref}   

\begin{document}

\newcommand{\ev}[0]{\mathbf{e}}
\newcommand{\cv}[0]{\mathbf{c}}
\newcommand{\fv}[0]{\mathbf{f}}
\newcommand{\Rv}[0]{\mathbf{R}}
\newcommand{\Tr}[0]{\mathrm{Tr}}
\newcommand{\ud}[0]{\uparrow\downarrow}
\newcommand{\du}[0]{\downarrow\uparrow}
\newcommand{\Uv}[0]{\mathbf{U}}
\newcommand{\Iv}[0]{\mathbf{I}}
\newcommand{\Hv}[0]{\mathbf{H}}
\newcommand{\kv}[0]{\mathbf{k}}
\newcommand{\qv}[0]{\mathbf{q}}

\setlength{\jot}{2mm}

\newcommand{\jav}[1]{{\color{red}#1}}
\newcommand{\cpm}[1]{{\color{blue}#1}}

\newcommand{\ds}[1]{{\color{blue}#1}}

\title{Emergent modular Luttinger liquid from spin-partitioned entanglement in the one-dimensional Hubbard model}

\author{\'Ad\'am B\'acsi}
\email{bacsi.adam@sze.hu}
\affiliation{Department of Mathematics and Physics, Sz\'echenyi Istv\'an University, 9026 Gy\H or, Hungary}
\affiliation{MTA-BME Lendület "Momentum" Open Quantum Systems Research Group, Institute of Physics, Budapest University of Technology and Economics, Műegyetem rkp. 3., H-1111, Budapest, Hungary}

\author{C\u at\u alin Pa\c scu Moca}
\affiliation{MTA-BME Lendület "Momentum" Open Quantum Systems Research Group, Institute of Physics, Budapest University of Technology and Economics, 
Műegyetem rkp. 3., H-1111, Budapest, Hungary}
\affiliation{Department of Theoretical Physics, Institute of Physics, Budapest University of Technology and Economics, M\H uegyetem rkp. 3., H-1111 Budapest, Hungary}
\affiliation{Department of Physics, University of Oradea,  410087, Oradea, Romania}

\author{Bal\'azs D\'ora}
\email{dora.balazs@ttk.bme.hu}
\affiliation{MTA-BME Lendület "Momentum" Open Quantum Systems Research Group, Institute of Physics, Budapest University of Technology and Economics, 
Műegyetem rkp. 3., H-1111, Budapest, Hungary}
\affiliation{Department of Theoretical Physics, Institute of Physics, Budapest University of Technology and Economics, M\H uegyetem rkp. 3., H-1111 Budapest, Hungary}

\begin{abstract}
 We study the spin-partitioned entanglement Hamiltonian of the one-dimensional repulsive Hubbard model. 
 By combining bosonization with exact diagonalization, we show that tracing out one spin species produces a \emph{modular Luttinger liquid}
 whose properties fundamentally differ from those of the physical system. While the modular spectrum is fully dispersionless and possesses a 
 momentum-independent entanglement gap, its eigenstates exhibit algebraic correlations governed by a single effective Luttinger parameter equal to the 
 geometric mean of the charge and spin Luttinger parameters. The resulting entanglement spectrum displays a universal branching hierarchy 
 in excellent agreement with exact diagonalization. We further demonstrate that the modular ground state is nearly identical to that of the spinless 
 Luttinger liquid. 
 These results uncover a universal modular structure in interacting one-dimensional fermionic systems.
\end{abstract}

\maketitle

\paragraph*{Introduction.\textemdash}
Quantum entanglement has emerged as a fundamental paradigm for characterizing interacting many-body 
systems~\cite{calabresecardy,eisert,amico,nielsen,horodecki,friis,Islam2015}, offering a unique window into 
universal properties and quantum correlations that are often inaccessible through conventional local 
observables. Beyond the bipartite entanglement entropy, the full structure of the reduced density 
matrix---encoded in the entanglement, or modular Hamiltonian~\cite{lihaldane,borchers,dalmonte}---contains 
detailed information about the underlying quantum state and provides a bridge~\cite{zeng2019} between quantum 
information theory and condensed-matter physics.

While most studies have focused on spatial bipartitions~\cite{srednicki}, partitioning the Hilbert space 
according to internal degrees of freedom provides a complementary perspective on quantum correlations. In 
itinerant fermionic systems without spin-orbit coupling, spin-partitioned entanglement isolates the nonlocal 
quantum correlations generated exclusively by electron-electron interactions. Since the ground state of a 
noninteracting system factorizes into independent spin-up and spin-down Fermi seas, the spin entanglement 
vanishes identically. Any finite spin entanglement therefore provides a direct measure of 
interaction-induced quantum correlations.

Spin-partitioned entanglement has attracted growing attention in settings ranging from Kondo impurities~\cite{Oh2006} 
to BCS superconductors~\cite{puspus,gao2007}, while recent quantum-gas experiments~\cite{bippus2026} 
have demonstrated direct access to spin entanglement in the Fermi-Hubbard model. Despite this progress, a 
fundamental question remains open: what is the universal low-energy entanglement Hamiltonian associated with 
a spin partition in an interacting one-dimensional metal? Since the low-energy physics of the Hubbard model 
is governed by a Luttinger liquid (LL) with spin-charge separation, it is natural to ask whether this 
universal structure survives in the modular Hamiltonian obtained after tracing out one spin species.

Here, we address this question by deriving the reduced density matrix and the universal spin-partitioned 
entanglement Hamiltonian of the one-dimensional repulsive Hubbard model. We show that the modular 
Hamiltonian realizes an unconventional hybrid structure, which we term a \emph{modular Luttinger liquid}. 
Although the physical system is gapless, the modular spectrum is fully dispersionless and exhibits a momentum-independent 
entanglement gap. Nevertheless, its eigenstates retain the algebraic correlations characteristic of a LL and 
are governed by a single effective LL parameter equal to the geometric mean of the charge and spin LL 
parameters. We confirm these analytical predictions numerically by demonstrating a universal branching 
hierarchy in the entanglement spectrum~\cite{lefevre} and by showing that the modular ground state is nearly 
identical to that of the spinless $t-V$ model.

\paragraph{Interacting electrons in one-dimension.\textemdash}

Our focus is on the spin entanglement Hamiltonian of the Hubbard model, but we start from the broader 
class of one-dimensional spinful interacting-electron systems described by Luttinger-liquid theory~\cite{giamarchi,nersesyan}. 
Consequently, the ensuing modular 
Hamiltonian is expected to apply to generic one-dimensional gapless spinful fermionic systems. 
The corresponding low-energy Hamiltonian is
 $H= H_\mathrm{kin} + H_\mathrm{int}$,
where the kinetic energy is bosonized as
 $H_\mathrm{kin} = \sum_{p\sigma}  v_{F}|p|   b_{p\sigma}^+ b_{p\sigma}$ 
with $v_F$ the Fermi velocity and $b_{p\sigma}$ the bosonic annihilation operator describing the elementary excitations in the spin-$\sigma$ sector with $\sigma=\uparrow$ or $\downarrow$. 
In the interaction, we distinguish between the forward scattering processes $g_2$ and $g_4$ as
\begin{gather}
H_\mathrm{int} =\sum_{p,\sigma}  g_4 |p| b_{p\sigma}^+ b_{p\sigma}  +\sum_{p} g_2 |p| \left(b_{p\uparrow}^+ b_{-p\downarrow}^+ +  b_{p\uparrow}b_{-p\downarrow}\right)\,.
\end{gather}
{Additional Umklapp and spin-backscattering terms are omitted because we focus on the regime with gapless spin and charge sectors. For the Hubbard model, this corresponds to $U>0$ and electron densities away from half filling.}
After introducing the conventional charge and spin fields~\cite{giamarchi}, the Hamiltonian ia 
rewritten in decoupled form
 $H= H_c + H_s$,
which is the manifestation of spin-charge separation for the low-energy physics of Luttinger liquids. 
This separation remains intact for all energies for the 
Hubbard model~\cite{essler2005}. These are written as
\begin{gather}
H_c=\sum_{p} \tilde{v}_c|p|   d_{pc}^+ d_{pc}, \hspace*{3mm} H_s=\sum_p \tilde{v}_s |p|  d_{ps}^+ d_{ps},
\label{eq:Hdiag}
\end{gather}
and are characterized by the LL parameters 
\begin{gather}
K_{c,s} = \sqrt{\frac{v_F \pm g_4 \mp g_2}{v_F \pm g_4 \pm g_2}},
\end{gather}
respectively, to be discussed later.
\paragraph{Reduced density matrix for the spin-$\uparrow$ sector.\textemdash}
We are now interested in the entanglement properties between the two spin species (spin-$\uparrow$ and spin-$\downarrow$) in the 
ground state of the Hamiltonian \eqref{eq:Hdiag}, $|\Psi_0\rangle$. The reduced density matrix is 
$\rho_\uparrow = \mathrm{Tr}_\downarrow\left[|\Psi_0\rangle\langle\Psi_0|\right]$, and the entanglement entropy follows from
 $S=-\mathrm{Tr}\left[ \rho_\uparrow \ln\rho_\uparrow\right]$. 
We use Ref.~\cite{Peschel2003}  to determine the structure of the reduced density matrix $\rho_\uparrow$. 
It is obtained as (see End Matter for the derivation)
\begin{gather}
\rho_\uparrow = \frac{\exp(-H_{ent})}{Z}= \prod_{p>0}\frac{e^{-H_{ent,p}}}{Z_p}
\label{eq:redm}
\end{gather}
where the entanglement Hamiltonian for a given pair of $(p,-p)$ modes is given by
\begin{gather}
H_{ent,p}=\varepsilon \left( b_{p\uparrow}^+ b_{p\uparrow} +
b_{-p\uparrow}^+ b_{-p\uparrow}\right) + \nonumber \\  \gamma\left(b_{p\uparrow}^+ b_{-p\uparrow}^+ + b_{-p\uparrow} b_{p\uparrow}\right) + \varepsilon-\beta,
\label{eq:redm0}
\end{gather}
where
\begin{gather}
\left(\begin{array}{c}
\varepsilon\\
\gamma
\end{array}\right)=\frac{\beta}{2}\left(\frac{1}{\sqrt{K_s K_c}}\pm \sqrt{K_c K_s}\right)
\end{gather}
are momentum-independent constants. Eq. \eqref{eq:redm0} is diagonalized as
\begin{gather}
H_{ent,p} = \beta\left(f_{p}^+ f_p + f_{-p}^{+}f_{-p}\right) 
\label{eq:Hent}
\end{gather}
and $Z_p = \left( 1-e^{-\beta} \right)^{-2}$ is the momentum independent normalization factor. Here, 
$f_p$ is a bosonic annihilation operator related to $b_{p\uparrow}$ and $b_{-p\uparrow}^+$ through a Bogoliubov transformation. The entanglement spectrum is 
\begin{gather}
\beta = 2\ln\left(\frac{\sqrt{K_s} + \sqrt{K_c}}{|\sqrt{K_s} - \sqrt{K_c|}}\right)
\label{eq:beta}
\end{gather}
{which takes the same value in every momentum channel. Consequently, the entanglement spectrum is identical in all momentum channels, and the eigenvalues of $H_{ent,p}$ are given by non-negative integer multiples of $\beta$, namely \{$0$, $\beta$, $2\beta$, $\dots$\}. Even though the original interacting Hamiltonian is gapless, the entanglement Hamiltonian for $\uparrow$ spins is gapped.}
Assuming $N$ boson modes, 
the bosonic vacuum, i.e. the zero energy state is
unique, the $\beta$ excitation energy corresponds to a single boson in one of the modes, therefore its degeneracy is $N$. The degeneracy 
of $2\beta$ excitation is $N+\binom{N}{2}$ as it can 
come from two bosons in the same mode or from two single boson excitations in distinct modes. The degeneracy of a  generic, 
$n\beta$ excitation scales to leading order as  $\binom{N}{n}$.

{Eq.~\eqref{eq:beta} is the static signature of spin-charge separation in this framework, governed entirely by the scaling dimensions and interaction strengths,}
which are parameterized by $K_{c,s}$. 
The eigenvalues of the entanglement spectrum~\cite{chandran} are driven entirely by how much the 
charge parameter $K_c$  differs from the spin parameter $K_s$. 
{While the differing spin and charge velocities drive the physical separation of wave packets in time, the differing Luttinger parameters provide the corresponding static imprint in the entanglement structure. Spin-partitioned entanglement can therefore be viewed as a measure of the ``distance'' between the decoupled charge and spin sectors.}

{With the explicit reduced density matrix for one spin species in hand, we can now analyze the properties of the entanglement Hamiltonian, also referred to as the modular Hamiltonian~\cite{bisognano,borchers,faulkner,Wen_2018}},
{defined by}
{
\begin{equation}
    H_{mod}=-\ln(\rho_\uparrow)=H_{ent}+\ln(Z).
    \label{eq:modham}
\end{equation}
}
{Its spectrum is completely flat and dispersionless, consistent with Ref.~\cite{lundgren2013}; therefore, the associated modular flow in fictitious time does not describe temporal propagation, and all modes oscillate with the same frequency.}
We clarify that the proper quantum mechanical time evolution of the system is still 
governed by the physical Hamiltonian, the modular flow in modular time $s$ is an auxiliary tool
to reveal the spectrum of the modular Hamiltonian. This is achieved with the knowledge of the reduced density matrix as $\exp(-iH_{ent} s)\sim \rho_\uparrow^{is}$.
The spectrum of $H_{ent}$ thus  represents   non-LL like behaviour.

{By contrast, its eigenfunctions remain fully LL-like, and the correlations decay algebraically in space. The ground-state wavefunction of $H_{ent}$}
is that of a spinless LL with a single LL parameter $\sqrt{K_cK_s}$, which is the 
geometric mean of the charge and spin LL parameters of the parent physical  Hamiltonian.
{After the Bogoliubov rotation from $b$ to $f$ bosons, this wavefunction takes a particularly simple form in Eq.~\eqref{eq:Hent}: it is the $f$-boson vacuum. The other eigenstates of $H_{ent}$ are excitations of this vacuum with increasing numbers of $f$ bosons. Via the inverse Bogoliubov rotation, they can be transformed back to the original $b$-boson language.}
Altogether, this modular LL, described by the entanglement Hamiltonian, is a non-CFT LL~\cite{giamarchi} with genuine power-law spatial 
correlations but nondispersive time dependence.

The entanglement entropy is calculated as
\begin{gather}
S = -\Tr\left[\rho_\uparrow\ln\rho_\uparrow\right]
= 2\sum_{p>0}\left[ \frac{\beta}{e^{\beta}-1} - \ln\left(1-e^{-\beta}\right)\right]
\label{eq:enteng1}
\end{gather}
where the sum over $p$ can be carried out due to the $p$-independence of $\beta$. This has been analyzed in related systems, i.e.  coupled 
LLs~\cite{furukawa,lundgren2013,lauchli}, therefore we refrain from its detailed analysis and only note that it satisfies a volume law and scales with the system size.


\paragraph{Hubbard model.\textemdash}

\begin{figure}[h]
\centering  
\includegraphics[width=0.8\columnwidth]{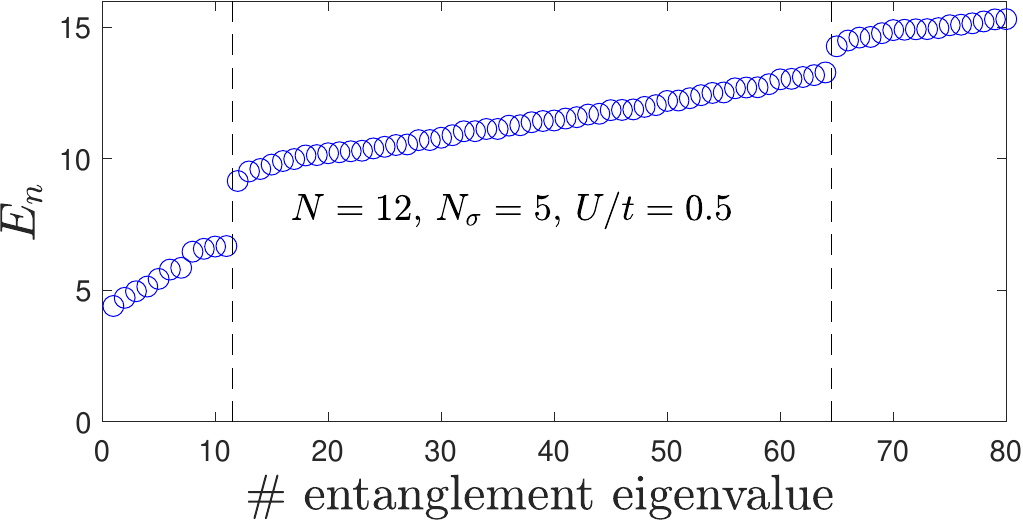}
\caption{Representative visualization of the first 80 eigenvalues of the entanglement Hamiltonian, displaying the branching property
of the spectrum. In particular, the first 11 and second 53 eigenvalues make the 1st and 2nd branch, respectively, separated by black dashed lines.}
\label{fig:enteigen}
\end{figure}

The  Hubbard model~\cite{essler2005,arovas} is defined as
\begin{gather}
H=-\sum_{j,\sigma} \frac{t}{2} \left(c_{j\sigma}^\dagger c_{j+1\sigma}+ \text{h.c.}\right)+U\sum_j n_{j\uparrow}n_{j\downarrow},
\label{hubbard}
\end{gather}
where $N$ sites are subject to periodic boundary conditions in one dimension and $n_{j\sigma}=c_{j\sigma}^\dagger c_{j\sigma}$.
The noninteracting dispersion is $\varepsilon_\sigma(k)=-t \cos(k)$ with the lattice constant set to unity. 
For the Hubbard model, when $g_2=g_4 = U/(2\pi)$, the Luttinger parameters are expressed in the weak coupling, small $U$ limit as
\begin{gather}
K_c \approx \frac{1}{\sqrt{1+\frac{U}{\pi v_F}}} \qquad K_s\approx \frac{1}{\sqrt{1-\frac{U}{\pi v_F}}},
\label{LLK}
\end{gather}
where $v_F=t\sin(k_F)$. Here we note that due to SU(2) spin invariance, $K_s=1$ is expected. 
However, this relationship holds for long chains (i.e. thermodynamic limit).
For short systems, {  $K_s$ is affected by logarithmic finite-size corrections generated by a marginally irrelevant operator}
and its behavior is characterized by a very slow, logarithmic drift toward exactly 
1 with the system size~\cite{Soffing_2013}.

\begin{figure}[h]   
\centering  
\includegraphics[width=0.9\columnwidth]{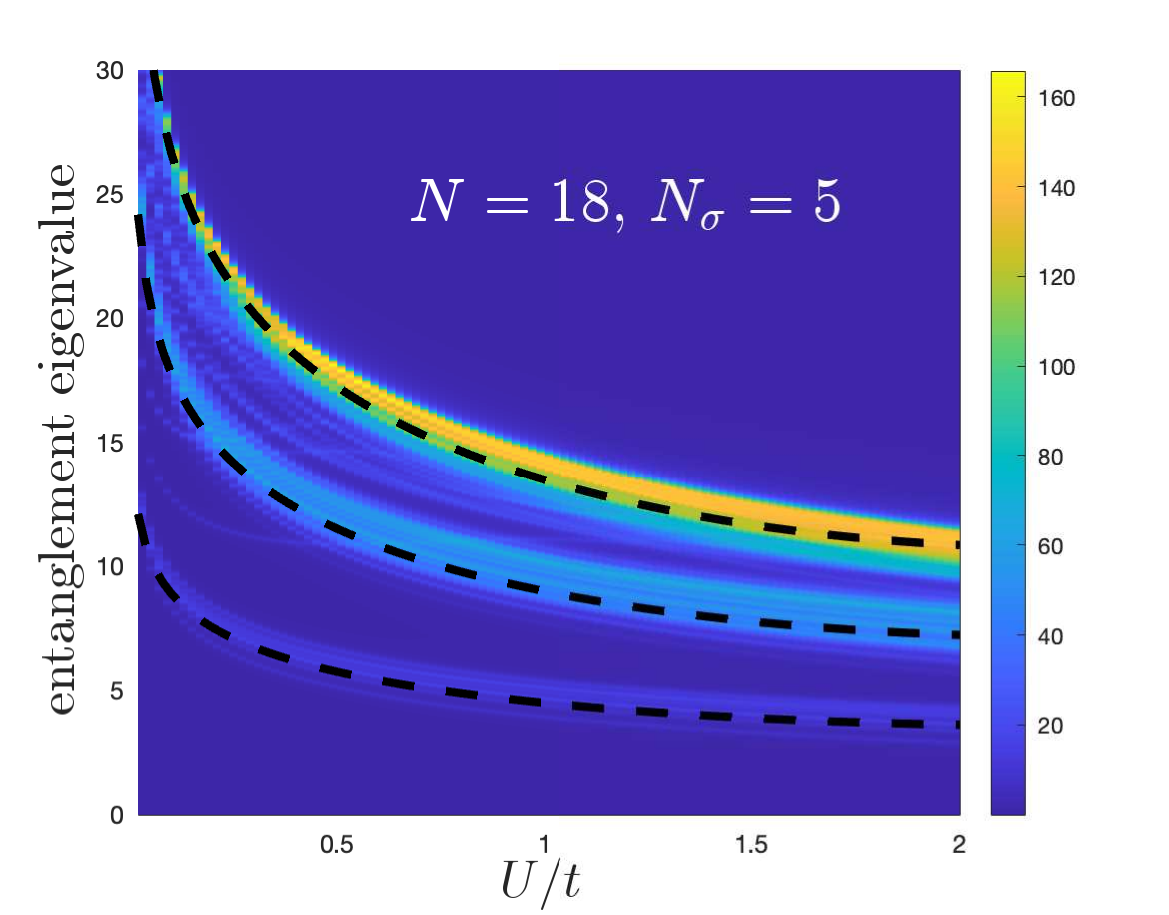}
\caption{Density plot of the evolution of the entanglement density of states, constructed from the first 400 largest eigenvalues {of the reduced density matrix} out of the possible $\binom{18}{5}$, 
revealing the structures in the entanglement eigenvalues. The dashed lines denote the bosonization results from Eq. \eqref{eq:beta}. The contribution of even 
larger eigenvalues can safely be neglected to the spin entanglement entropy.}
\label{fig:spinent}
\end{figure}
\paragraph{Entanglement spectrum.\textemdash}
We study numerically the spin entanglement of the Hubbard model \eqref{hubbard} using many-body exact diagonalization (ED).
 After finding the ground state of the interacting system, the down spin electrons are traced
out to get to the reduced density matrix of the up spin electrons, $\rho_\uparrow$, which describes the entanglement Hamiltonian through Eq. \eqref{eq:redm}.
Then, $\rho_\uparrow$ is diagonalized to obtain the spectrum of $H_{ent}$, denoted by $E_n$.
The numerically obtained entanglement spectrum is plotted in Figs.~\ref{fig:enteigen} and~\ref{fig:spinent}.
In particular, we plot the entanglement density of states~\cite{geraedts} in Fig.~\ref{fig:spinent},
which is defined from the spectrum of the entanglement Hamiltonian as
\begin{gather}
g(E)=\sum_n\delta(E-E_n),
\end{gather}
similarly to the conventional density of states of electrons~\cite{abrikosov}.

\begin{figure}[h]
\centering
\includegraphics[width=0.95\columnwidth]{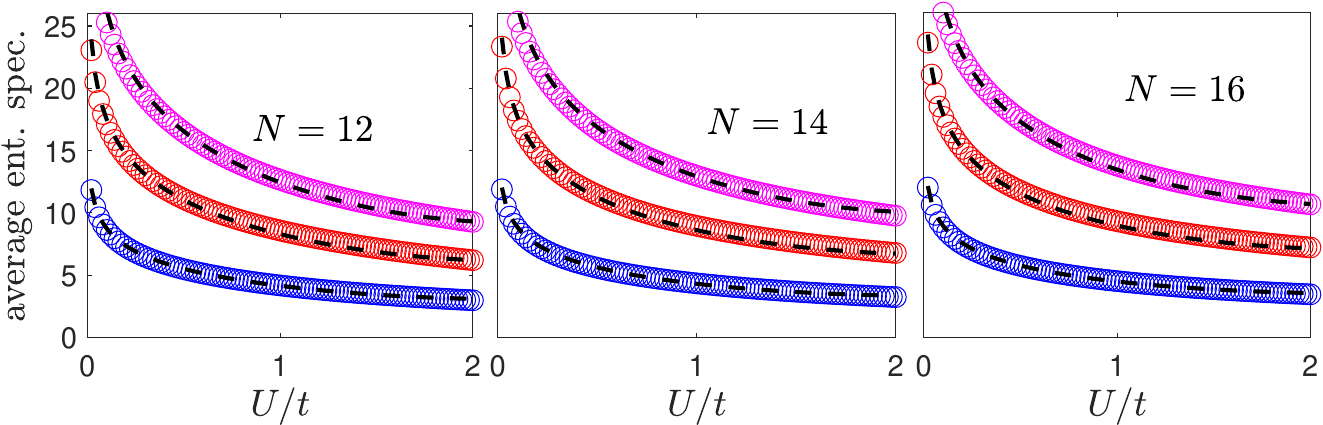}
\caption{The \emph{average} entanglement eigenvalue of the first 3 lowest branches are plotted for 
$N_\uparrow=N_\downarrow=5$ particles with increasing system size. The black dashed
lines denote the bosonization results from Eq.~\eqref{eq:beta}, predicting $\beta\sim \ln(t/U)$ in the weak coupling limit.}
\label{fig:entspec4n}
\end{figure}

Using the branching property of the spectrum, we calculate the average
entanglement eigenvalue within a given branch by evaluating the arithmetic mean of eigenvalues within a given branch 
and compare that to bosonization prediction in Fig.~\ref{fig:entspec4n}, displaying excellent agreement.

In order to achieve a meaningful comparison between numerics and bosonization,
we need to take into account that $K_s\neq 1$ due to the short chains considered. Therefore, we use the expression~\cite{Sano,Soffing_2013}
\begin{gather}
1/K_s^2\approx 1-a_1\frac{U}{\pi t \sin(k_F)}+a_2\frac{2U^2}{\pi t^2\sin(k_F)}+\dots,
\label{Ks}
\end{gather}
where $k_F=\pi f$ and $f=N_{\uparrow}/N$ is the filling fraction and $a_{1,2}$ are fitting parameters, which can depend on the system size and filling. 
Our results are summarized in Table ~\ref{table1} by comparison with the numerical data in Fig.~\ref{fig:entspec4n}.
{These parameters steadily decrease with system size, indicating that $K_s=1$ is approached only very slowly in the thermodynamic limit.}
We could also consider higher order corrections to $K_c$ in Eq.~\eqref{LLK}, similarly to $K_s$, but this would only increase the number of fitting parameters and not the quality of the fit,
therefore we refrain from considering higher order terms and use Eq.~\eqref{LLK} for $K_c$.

\begin{table}[t]
\centering
\renewcommand{\arraystretch}{1.25}
\setlength{\tabcolsep}{8pt}
\begin{tabular}{c|cccc}
\hline\hline
System size $N$ & 12 & 14 & 16 & 18 \\
\hline
$a_1$ & 2.06 & 1.77 & 1.44 & 1.28 \\
$a_2$ & 0.25 & 0.22 & 0.18 & 0.17 \\
\hline\hline
\end{tabular}
\caption{System-size dependence of the fitting parameters entering $K_s$ in Eq.~\eqref{Ks}.}
\label{table1}
\end{table}

In the non-interacting case, $U=0$, the ground state wavefunction is a product of two Slater determinants for the two spin species, therefore there is no entanglement 
between the different spin directions in accordance with the physical expectations. This is also reflected in $\beta\rightarrow\infty$ from Eq.~\eqref{eq:beta}, where both $K_s=K_c=1$ in the
non-interacting limit.
{With increasing interaction strength, the entanglement increases monotonically and, for small $U$, grows as $S\propto -U^2\ln(U)$.}

Finally, we note that in one dimension, a general relation between entanglement and particle number fluctuations~\cite{song} has been put forward for spatial bipartitioning. 
In our case with spin bipartitioning,
the total number of up and down spins is conserved separately, therefore their fluctuations are zero over the whole system, while the corresponding spin partitioned entanglement is finite.
Therefore, the relation between spin partitioned entanglement and spin fluctuations does not hold.

\paragraph{Entanglement wavefunction.\textemdash}

So far, we focused on the universal features of the entanglement spectrum. However, the eigenfunctions of the entanglement Hamiltonian 
are also relevant~\cite{zhu2019,toldin} for a complete characterization of $H_{ent}$ and its ground state wavefunction corresponds to a single component LL. We investigate this 
by focusing on the ground state wavefunction of one-dimensional spinless electrons within the $t-V$ model, given by
\begin{gather}
H=-\sum_{j} \frac{t}{2} \left(c_{j}^\dagger c_{j+1}+ \text{h.c.}\right)+V\sum_j n_{j}n_{j+1},
\label{xxz}
\end{gather}
which consists of $N$ sites and the number of spinless particles is $N_\uparrow$.
This model  is equivalent to the XXZ Heisenberg model after a Jordan-Wigner transformation~\cite{giamarchi,nersesyan} and is characterized by a single LL parameter $K_V$.
For $V>0$ and away from half filling, this model realizes a single component LL~\cite{giamarchi}. We numerically 
diagonalize this model using many-body ED for the same filling $f$
as the Hubbard model and search for the maximum overlap of the ground state wavefunction of Eq. \eqref{xxz} 
with the ground state wavefunction of the entanglement Hamiltonian for a given $U$ in the Hubbard model
by varying $V$. In a way, the ground state wavefunction of Eq.~\eqref{xxz} can be thought of as a variational wavefunction with variational parameter  $V$ and 
we want to maximize its overlap with the ground state wavefunction of the entanglement Hamiltonian with fixed $U$.

\begin{figure}[h!]
\centering
\includegraphics[width=0.8\columnwidth]{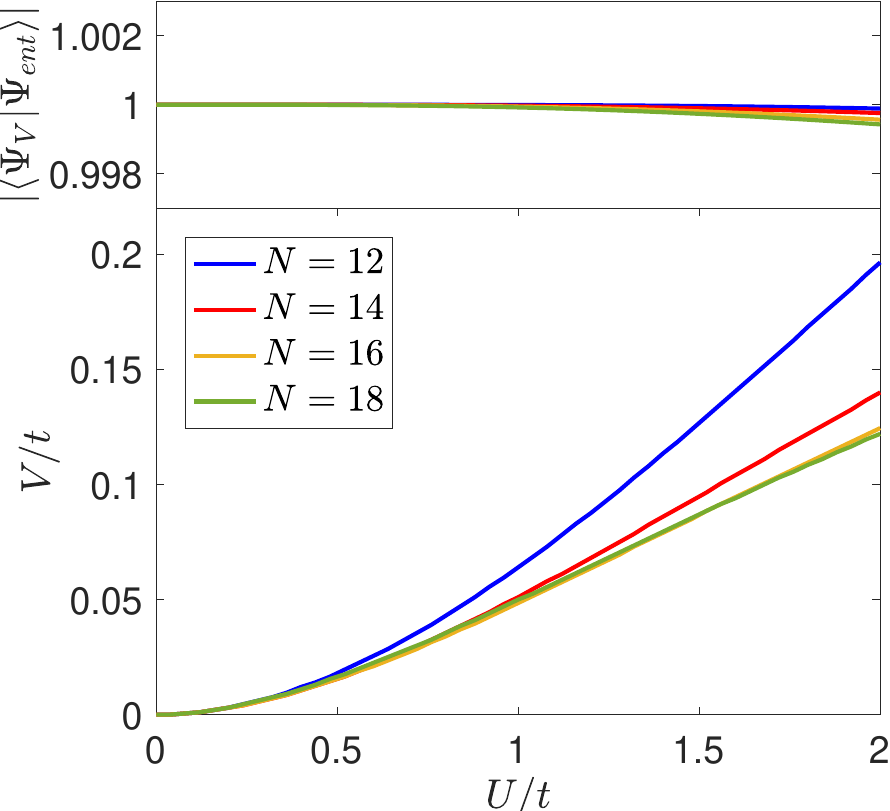}
\caption{The overlap of the ground state LL wavefunction of Eq.~\eqref{xxz}, $|\Psi_V\rangle$,
 and the ground state wavefunction of the spin entanglement Hamiltonian of the Hubbard model, $|\Psi_{ent}\rangle$,
is shown from many-body ED in the upper panel, for the $V-U$ combinations in the lower panel. Here, $N_\uparrow=N_\downarrow=5$, and this is also the number of particles in the $t-V$
model.
The two wavefunctions for several system sizes are practically equivalent to each other.
}
\label{fig:couplingoverlap}
\end{figure}

In Fig.~\ref{fig:couplingoverlap}, we display the results of this comparison. 
{The ground-state wavefunction of Eq.~\eqref{xxz} exhibits a very large, nearly perfect overlap with the ground-state wavefunction of the spin entanglement Hamiltonian of the Hubbard model for the corresponding $V$-$U$ pairs, even for the finite-size systems considered here. These values are consistent with $K_V=\sqrt{K_cK_s}$. Even for relatively strong Hubbard interactions, $U>t$, the effective single-component LL remains only weakly interacting, $V<t$, while still being clearly distinct from a noninteracting system.}

\paragraph{Experimental possibilities.\textemdash}
{Beyond condensed-matter realizations~\cite{giamarchi,nersesyan}, the Hubbard model has been implemented with cold atoms~\cite{tarruell2018} and in quantum simulations on superconducting qubits~\cite{alam,hartnett,arute}. In such platforms, the two spin species can be spatially separated to form a two-leg ladder, with one spin species on each leg. Hopping then occurs along the legs, while the Hubbard interaction acts along the rungs, together with single-particle hopping between the legs. In this language, spin-partitioned entanglement can be rephrased as a spatial bipartition with a cut along the rungs, so that the spin entanglement becomes equivalent to the spatial entanglement between the two legs of the ladder~\cite{Islam2015}. In addition, spectroscopy of the entanglement Hamiltonian may also be feasible~\cite{delmonte2018}.}

\paragraph{Conclusions.\textemdash}

We have derived the universal spin-partitioned entanglement Hamiltonian of the one-dimensional repulsive Hubbard model by combining
bosonization with exact diagonalization. The resulting modular Hamiltonian realizes an unconventional Luttinger liquid: although its
entanglement spectrum is completely dispersionless and separated by a momentum-independent entanglement gap, its eigenstates retain
the algebraic correlations of a single component Luttinger liquid.  
The predicted 
universal branching structure of the entanglement spectrum is quantitatively confirmed by exact diagonalization, and the modular ground
state exhibits an almost perfect overlap with the ground state of the spinless t–V model.

Our work demonstrates that internal symmetry partitions can support universal modular theories distinct
from the physical low-energy dynamics while preserving their critical correlations. This framework opens the door to exploring modular
Hamiltonians in more general one-dimensional correlated systems, including superconducting, topological, and non-Hermitian phases, where
the interplay between interactions and internal degrees of freedom may reveal new universal entanglement structures.

\begin{acknowledgments}
This work was supported by the National Research, Development and Innovation Office - NKFIH  Project No. K142179,
by a grant of the Ministry of Research, Innovation and
 Digitization, CNCS/CCCDI-UEFISCDI, under projects number
PN-IV-P1-PCE-2023-0159 and PN-IV-P1-PCE-2023-0987.
This work was also supported by the HUN-REN Hungarian Research Network through the Supported Research Groups
Programme, HUN-REN-BME-BCE Quantum Technology Research Group (TKCS-2024/34).
 This work was performed in part at the Aspen Center for Physics, which is supported by a grant from the Simons Foundation (1161654, Troyer).
\end{acknowledgments}

\bibliographystyle{apsrev}
\bibliography{wboson1,refgraph}
\appendix 
\section{End matter}
\paragraph{Charge and spin separation in the Luttinger liquid.\textendash}
\label{spinchargeham}
{We introduce the bosonic operators corresponding to the charge and spin sectors as}
\begin{gather}
b_{pc} = \frac{b_{p\uparrow} + b_{p\downarrow}}{\sqrt{2}} \qquad\qquad b_{ps} = \frac{b_{p\uparrow} - b_{p\downarrow}}{\sqrt{2}} \,,
\end{gather}
{Consequently, the charge-sector Hamiltonian takes the form}
\begin{gather}
H_c = \sum_{p} \frac{\tilde{v}_c |p|}{2} \left[ \left(\frac{1}{K_c} + K_c\right)  b_{pc}^+ b_{pc} +
\right. \nonumber \\ + \left. \left(\frac{1}{K_c} - K_c\right) \frac{b_{pc}^+ b_{-pc}^+ + b_{-pc} b_{pc} }{2}\right]
\end{gather}  
where $\tilde{v}_c=\sqrt{\left(v_F+ g_4\right)^2 - g_2^2}$ is the velocity of elementary excitations and
$K_c = \sqrt{\frac{v_F + g_4 - g_2}{v_F + g_4 + g_2}}$ is the Luttinger parameter
which fulfills $K_c<1$ in the case of repulsive interaction $g_2>0$ and $g_4>0$.
Similarly, the Hamiltonian of the spin sector is reformulated as
\begin{gather}
H_s = \sum_{p} \frac{\tilde{v}_s |p|}{2} \left[ \left(K_s + \frac{1}{K_s} \right) b_{ps}^+ b_{ps}
\right. - \nonumber \\ - \left. \left(K_s - \frac{1}{K_s} \right) \frac{ b_{ps}^+ b_{-ps}^+ + b_{-ps} b_{ps}}{2}\right]
\end{gather}  
with
\begin{gather}
\tilde{v}_s=\sqrt{\left(v_F- g_4\right)^2 - g_2^2} \qquad K_s = \sqrt{\frac{v_F - g_4 + g_2}{v_F - g_4 - g_2}}.
\end{gather}  

By applying the Bogoliubov transformation $b_{pc/s} = u_{+,c/s} d_{pc/s} - u_{-,c/s} d_{-pc/s}^+$
with the coefficients $u_{\pm,c/s} = \frac{1}{2}\left(1/\sqrt{K_{c/s}} \pm \sqrt{K_{c/s}}\right)$,
the Hamiltonian is diagonalized.

\paragraph{Calculation of the reduced density matrix.\textemdash}
\label{sec:peschel}
In this section, we present the derivation of the reduced density matrix by following Refs.~\cite{Peschel2003} and~\cite{lundgren2013}. The key idea is to define a 
Gaussian-structured density matrix such that the following two-point correlation functions are captured.
\begin{gather}
\langle \Psi_0 | b_{p\uparrow}^+ b_{p\uparrow}|\Psi_0\rangle = 
\frac{1}{8}\left(\frac{1}{K_c}+K_c + \frac{1}{K_s} + K_s - 4\right) \nonumber \\
\langle \Psi_0|b_{p\uparrow}^+ b_{-p\uparrow}^+ |\Psi_0\rangle = 
-\frac{1}{8}\left(\frac{1}{K_c} - K_c + \frac{1}{K_s} - K_s\right)
\label{eq:2pcorr}
\end{gather}
where $|\Psi_0\rangle$ is the ground state of the Hamiltonian given in Eq. \eqref{eq:Hdiag} of the main text.
{Since the bosonized theory is quadratic, the reduced density matrix is quadratic as well. Its generic form is}
\begin{gather}
\rho_\uparrow = \prod_{p>0}\mathcal{N}_p e^{-\Big(\varepsilon_p\left( b_{p\uparrow}^+ b_{p\uparrow} + 
b_{-p\uparrow}^+ b_{-p\uparrow}\right) + \gamma_p\left(b_{p\uparrow}^+ b_{-p\uparrow}^+ + b_{-p\uparrow} b_{p\uparrow}\right)\Big)}
\end{gather}
with some parameters $\varepsilon_p$ and $\gamma_p$ and with the normalization factor $\mathcal{N}_p$.  
The exponent is diagonalized by the Bogoliubov transformation
$b_{p\uparrow} = \mu_{+,p} f_p - \mu_{-,p} f_{-p}^+$
with
$\mu_{\pm,p} =\frac{1}{\sqrt{2}} \sqrt{\frac{\varepsilon_p}{\beta_p}\pm 1}$
leading to
\begin{gather}
\rho_\uparrow = \prod_{p>0}\frac{e^{-H_{ent,p}}}{Z_p}  \\
H_{ent,p} = \beta_p\left(f_{p}^+ f_p + f_{-p}^{+}f_{-p}\right) \\
 Z_p = \left( 1-e^{-\beta_p} \right)^{-2}=\frac{e^{\beta_p-\varepsilon_p}}{\mathcal{N}_p}, 
\label{eq:Hent2}
\end{gather}
where $\beta_p = \sqrt{\varepsilon_p^2 - \gamma_p^2}$. The entanglement Hamiltonian of the form \eqref{eq:Hent2} shows that the entanglement is determined by the parameter $\beta_p$ only.

By using the reduced density matrix, the relevant two-point correlation functions are calculated as
\begin{gather}
\Tr\left[ \rho_\uparrow b_{p\uparrow}^+ b_{p\uparrow}\right] = \frac{\varepsilon_p}{2\beta_p}\mathrm{coth}\left(\frac{\beta_p}{2}\right) - \frac{1}{2}  \nonumber \\
\Tr\left[ \rho_\uparrow b_{p\uparrow}^+ b_{-p\uparrow}^+\right] = -\frac{\gamma_p}{2\beta_p}\mathrm{coth}\left(\frac{\beta_p}{2}\right)
\end{gather}
which must coincide with Eqs. \eqref{eq:2pcorr}. The equality condition results in
\begin{gather}
\beta_p = 2\ln\left(\frac{\sqrt{K_s} + \sqrt{K_c}}{|\sqrt{K_s} - \sqrt{K_c}|}\right):=\beta \,.
\end{gather}
We note that the higher order correlation functions are also correctly reproduced by the reduced density matrix due to Wick's theorem \cite{Peschel2003}. 

In the weak coupling limit, $\varepsilon\sim\ln(t/U)$ while $\gamma\sim U\ln(t/U)$. Consequently, $\beta\sim\ln(t/U)$, and  $H_{ent}$, though possesses momentum
independent terms,  describes a weakly interacting theory since $|\gamma|\ll |\varepsilon|$ in the weak coupling limit.


\end{document}